\documentclass[a4paper,11pt]{article}
\pdfoutput=1
\usepackage{jheppub,relsize}
\usepackage[T1]{fontenc} 
\usepackage{amsmath,amssymb,graphicx,bbm,mathrsfs,nicefrac}
\usepackage{tikz}
\usepackage{amsmath}
\usepackage{amssymb}
\usepackage{graphicx,textcomp,float,gensymb,wrapfig, enumitem,comment,dsfont,framed,slashed,appendix,wrapfig,xcolor}
\usepackage{ bbold }
\usepackage{braket} 
\usepackage{ mathrsfs }
\usepackage[small]{caption}
\usepackage{subcaption}
\usepackage{etoolbox,hyperref,makecell}
\usepackage{cancel}
\usepackage{feynmp}
\usetikzlibrary{arrows}
\usepackage{empheq}
\patchcmd{\abstract}{\null\vfil}{}{}{}
\newcommand{\be}{\begin{equation}} 
\newcommand{\ee}{\end{equation}}  
\newcommand{\bea}{\begin{eqnarray}}  
\newcommand{\eea}{\end{eqnarray}}

\usepackage{enumitem}
\makeatletter
\newcommand{\mylabel}[2]{#2\def\@currentlabel{#2}\label{#1}}
\makeatother

\begin{document}%in the revtex template, the title and author info should be included after the begin doc command.
\title{Dark matter amnesia in out-of-equilibrium scenarios 
%via a Dark Photon Portal
}
\author[a]{Joshua Berger,}
\author[b]{Djuna Croon,}
\author[c,d]{Sonia El Hedri,}
\author[e]{Karsten Jedamzik,}  
\author[f]{Ashley Perko,}
\author[f]{Devin G.~E.~Walker}
  
  \affiliation[a]{Department of Physics and Astronomy, University of Pittsburgh, Pittsburgh, PA 15260, USA}
  \affiliation[b]{TRIUMF Theory Group, 4004 Wesbrook Mall, Vancouver, B.C. V6T2A3, Canada}
  \affiliation[c]{Nikhef, Theory Group, Science Park 105, 1098 XG, Amsterdam, The Netherlands}
  \affiliation[d]{Laboratoire Leprince Ringuet, \'Ecole Polytechnique, 91120 Palaiseau, France}
  \affiliation[e]{Laboratoire Univers et Particules de Montpellier, UMR5299-CNRS, Universite Montpellier II, 34095 Montpellier, France}
  \affiliation[f]{Department of Physics and Astronomy, Dartmouth College,
  Hanover, NH 03755, USA}
  
  \emailAdd{josh.berger@pitt.edu}
  \emailAdd{dcroon@triumf.ca}
  \emailAdd{elhedrisonia@gmail.com}
  \emailAdd{karsten.jedamzik@univ-montp2.fr}
  \emailAdd{perko@dartmouth.edu}
  \emailAdd{devin.g.walker@dartmouth.edu}
  
  \preprint{PITT-PACC-1822}

\date{\today}

\abstract{
    Models in which the dark matter is produced at extremely low rates from the annihilation of Standard Model particles in the early Universe allow us to explain the current dark matter relic density while easily evading the traditional experimental constraints. In scenarios where the dark matter interacts with the Standard Model via a new physics mediator, the early Universe dynamics of the dark sector can be particularly complex, as the dark matter and the mediator could be in thermal and chemical equilibrium with each other. This equilibration takes place via number-changing processes such as double Compton scattering and bremsstrahlung, whose amplitudes are cumbersome to calculate. In this paper, we show that in large regions of the parameter space, these equilibration mechanisms do not significantly affect the final dark matter relic density. In particular, for a model with a light dark photon mediator, the relic density can be reasonably estimated by considering that the dark matter is solely produced through the annihilation of Standard Model particles. This result considerably simplifies the treatment of a large class of dark matter theories, facilitating in particular the superimposition of the relic density constraints on the current and future experimental bounds.  
}

\maketitle
\flushbottom

\section{Introduction}
\label{sec:introduction}
Understanding the nature of dark matter (DM) is one of the most pressing unresolved problems in particle physics and cosmology. An important class of dark matter theories are models where the dark sector has a gauge structure, the simplest of these scenarios being extensions of  the Standard Model (SM) by a dark $U(1)'$ gauge group. This symmetry is associated with a dark photon $A'$ and can possibly be spontaneously broken by a new Higgs boson $\Phi$. In the general case, $A'$ and $\Phi$ can mix with the SM hypercharge gauge boson and with the SM Higgs respectively, thereby connecting the dark and visible sectors~\cite{Holdom:1985ag,DelAguila:1993px,DelAguila:1995fa,Babu:1996vt,Babu:1997st}. In these so-called ``portal models'', the dark matter can be produced from the annihilation of Standard Model and other dark sector particles in the early Universe, and can also annihilate into these particles at later times. This finding has led to the hypothesis that these production and annihilation processes completely determine the final dark matter relic density. Under this hypothesis, the couplings and the masses of the dark sector particles would be strongly tied to this relic density, thus allowing for a determination of how the viable regions of the parameter space of existing models intersect with the current and future experimental constraints. 

Linking the relic density requirements for gauge and Higgs portal models to experimental constraints is all the more important, as the corresponding signatures are extremely diverse. One particularly interesting avenue is to search not directly for the dark matter, but for the mediator itself. New dark gauge bosons, in particular, could be directly produced at colliders, affect the flavor observables, or be produced at beam dump experiments~\cite{Englert:2013gz,Kumar:2012ww,Englert:2012ha,LopezHonorez:2012kv,Lebedev:2012zw,Batell:2011pz,Djouadi:2011aa,Englert:2011aa,Baek:2011aa,Lebedev:2011iq,Brivio:2015kia,Sun:2015oea,Freitas:2015hsa,Fedderke:2015txa,Khoze:2015sra,Bishara:2015cha,Chao:2015uoa,Falkowski:2015iwa,Chao:2014ina,Chacko:2013lna,Choi:2013qra}. More recently, however, indirect astrophysical and cosmological probes have been discussed for models where the mediators are particularly light or couple extremely weakly to the SM.  For particularly weak coupling, the dark mediators are sufficiently long lived as to decay around the era of Big Bang Nucleosynthesis (BBN) or recombination \cite{Berger:2016vxi,Fradette:2015nna}. In this case, they are constrained by the observed light element yields in the Universe and the success of the $\Lambda$CDM cosmology at predicting the Cosmic Microwave Background (CMB) spectrum respectively.  The sensitivity to CMB distortions will be greatly enhanced by a possibly upcoming PIXIE experiment~\cite{Kogut:2011xw}.

Constraints from direct and indirect detection are also possible.  Although the coupling of the dark matter to the SM is very weak, the mediator can be rather light in these scenarios and spin independent direct detection constraints are strong~\cite{Aprile:2018dbl,Agnes:2018ves,Ren:2018gyx,Cui:2017nnn,Petricca:2017zdp,Akerib:2016vxi}. On the other hand, the lightness of the mediator works against the constraints on these models as the recoil momentum transfer becomes comparable to the mediator mass~\cite{Kaplinghat:2013yxa}. The lightness of the mediator limits the constraints to the point where they are generally irrelevant to the models considered below. Indirect detection does not face this issue, but the dominant annihilation channel for dark matter coupling to a light mediator is into mediators, rather than SM particles, softening the observed spectrum and weakening the constraints somewhat~\cite{Elor:2015bho}. The annihilation cross-section into mediators is rather large, though, leading to some significant constraints if the dark matter coupling to the mediator makes up the totality of the observed dark matter abundance in the universe.

One crucial issue with cosmological experimental probes is that they constrain regions of the parameter space where the couplings of the dark matter to the SM are extremely small. In these regions, the dark matter can never be in equilibrium with the SM and thermal dark matter models will not be viable. Understanding how cosmological constraints on light gauge and Higgs bosons affect dark matter models therefore seems to require a thorough understanding of the dynamics of the dark matter when it is produced out of equilibrium. The associated scenarios can be extremely complex as the internal dynamics of the dark sector could significantly affect the dark matter evolution. Taken at face value, resultant dark matter densities could differ by orders of magnitude depending on the degree of dark sector equilibration.

To illustrate this, let us consider two extreme cases: (a) full dark sector equilibration and (b) complete absence of interactions between particles in the dark sector. In either case, dark sector particles are produced via freeze-in~\cite{Hall:2009bx} of the mediator $A^{\prime}$ and the dark matter $\chi$ particles (having $U(1)'$ charges), with the former usually produced in much larger numbers. However, whereas in case (b) the dark matter abundance is simply given by its freeze-in value $n^{\rm fi}_{\chi}$, in case (a) production of dark sector particles via $2\to 3$ processes and pair production of $\chi'$s via $A'$ annihilation lead to parametrically larger $\chi$ and $A'$ abundances.  Here $2\to 3$ processes favor the production of extra dark sector particles since the typical energy of a dark sector particle is approximately $T$, the visible sector temperature, which is far too large for an equilibrated dark sector at temperature $T'\ll T$. Thus extra dark particles have to be produced for equilibration to occur. A rough estimate of the equilibrated $\chi$ abundance can be obtained via the energy leaked into the dark sector, i.e. $T n^{\rm fi}_{\chi}\sim T'^4$ and $n^{\rm eq}_{\chi} \sim 
T'^3$, leading
to
\be 
\frac{n^{\rm eq}_{\chi}}{n^{\rm fi}_{\chi}} 
\sim  \left( \frac{T^3}{n^{\rm fi}_{\chi}} \right)^{1/4} \ , 
\ee 
which is of the order $10^2-10^3$ in the parts of parameter space where the relic abundance saturates $\Omega_\text{DM}$. Note that this ratio represents a serious underestimate as the bulk of the visible energy is expected to be leaked into the mediator, an effect not taken into account in the estimate. For such scenarios, the relic abundance calculation is thus plagued by large uncertainties. Note that such scenarios have been studied in Ref.~\cite{Chu:2011be}, and though this work has outlined important features of dark matter
portal scenarios, it has not properly addressed these equilibration uncertainties.

This paper will show that a class of dark sector models which are distinguished by a rich thermal evolution are degenerate from a phenomenological perspective. In particular, we will demonstrate that, in large regions of the parameter space, the dark matter relic density depends only weakly on the details of the chemical equilibration of the dark sector in the early Universe. This result is due to the fact that, in general, the dark matter produced via the annihilation of the dark mediator will have enough time to fully annihilate before the dark sector freezes out. Hence, the dark matter relic density in today's Universe will be fully determined by its production through the annihilation of SM particles and possibly also by its late time annihilation into dark sector particles. Since the cross-section for the latter process is nearly constant at low temperature this annihilation rate depends only weakly on the degree of equilibration in the dark sector. For wide ranges of parameters, it is therefore possible to make order-of-magnitude estimates of the dark matter relic density without any knowledge of the magnitude of the number-changing processes involving dark sector particles. This result thus allows us to quickly and straightforwardly delineate the regions of parameter space most relevant in the
quest of dark matter.

The rest of this paper is outlined as follows. In section~\ref{sec:scenarios} we introduce the toy model that we use for our study, which is an extension of the SM with a fermionic dark matter singlet and a new $U(1)'$ symmetry associated with a dark photon. We then describe the two extreme scenarios that we are going to consider: the case of full equilibrium within the dark sector, where the dark matter can be produced from dark photon annihilation, and the case where the DM and the dark photon never equilibrate. In section~\ref{sec:numerical}, we detail the procedure we used to scan the parameter space for our example model. We then show that, when the dark photon is much lighter than the dark matter both scenarios lead to similar relic densities. Finally, we conclude in section~\ref{sec:discussion} that for a majority of the parameter space of interest, the dark matter relic abundance can be determined using the simplified scenarios discussed in section~\ref{sec:scenarios} rather than the full out-of-equillibrium calculation.

\section{Model and scenarios}
\label{sec:scenarios}
\begin{figure}
    \centering
    \includegraphics[width=0.6\linewidth]{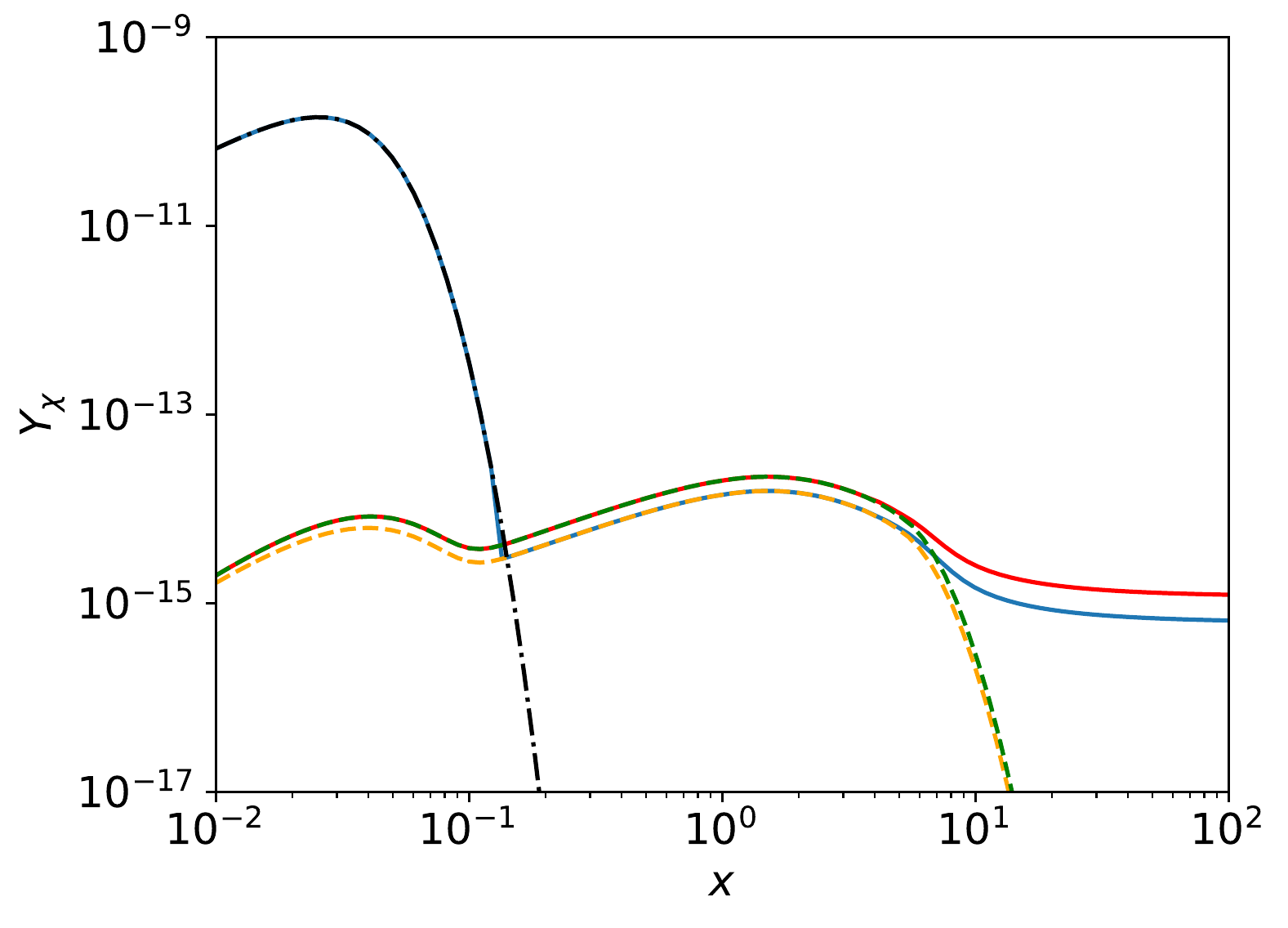}
    \caption{\label{fig:1Drelic} Comoving number density of the dark matter $Y_\chi$ as a function of $x = m_\chi/T$ for $m_\chi = 800$~MeV, $m_{A'} = 20$~MeV, $\epsilon = 4.3\times 10^{-12}$, and $\alpha' = 0.01$. The blue and red solid lines represent the values of $Y_\chi$ in scenarios~\ref{scenario1a} and \ref{scenario2b} respectively. The orange and green dashed lines show $Y_{QSE}$ (defined in equation~\ref{eq:yqse}) in the following two limit cases: the dark sector has an equilibrium temperature $T'$ and the dark sector has zero temperature. Finally, the black dot-dashed line shows the value of $Y_\chi$ when the dark sector is at equilibrium at a temperature $T'$. Here, we started integrating the evolution equations only when the DM departs from equilibrium and, before that time, approximated its density by its value in its various equilibrium states, hence the sharp turn for the blue curve around $x = 0.15$.}
\end{figure}

We consider a simple extension of the Standard Model with a broken $U(1)'$ gauge group associated with a dark photon $A'$, and a fermionic dark matter candidate $\chi$. The new $U(1)'$ gauge group kinetically mixes with hypercharge with strength $\epsilon$. The model is described by the following Lagrangian:
\bea \mathcal{L} &=& - \frac{1}{4} \hat{B}_{\mu \nu} \hat{B}^{\mu \nu} - \frac{1}{4} \hat{F'}_{\mu \nu} \hat{F'}^{\mu \nu} -  \frac{\epsilon}{2} \hat{B}_{\mu \nu} \hat{F'}^{\mu \nu} \\
&&+ \bar{\chi} \left[ \gamma^\mu (i \partial_\mu - g' A'_\mu) - m_\chi  \right] \chi + \frac{1}{2} m_{A'}^2 (A'_\mu)^2 \ .
\eea 
Here, we do not introduce any dark scalars and hence we give $A'$ a Stueckelberg mass. In what follows, we focus on the regime where the dark photon is light but can still decay to SM particles, that is, in the mass range $2m_e < m_{A'} <1 \ \text{GeV}$. 
%%%
The condition for thermal equilibrium between the dark sector and the Standard Model is roughly given by\footnote{We expect other contributing processes, such as production of $A'$ by inverse decay, to be subdominant.} 
        \be \langle n_f \sigma_{ff\rightarrow\chi\chi} v \rangle \gtrsim H  \ , \label{eq:equilibrium}\ee
where $n_f$ is the SM fermion density and where $\langle \sigma_{ff\rightarrow\chi\chi} v \rangle$ is the velocity-averaged cross-section for the processes connecting the DM to the SM, and is proportional to $\epsilon^2$.
%%%
In this paper we consider only very small values for $\epsilon$, in the range $10^{-15} < \epsilon < 10^{-9}$, for which condition~\ref{eq:equilibrium} is not satisfied. In this regime, instead of reaching its equilibrium value before freezing out at later times, the DM abundance initially grows at a very slow rate~\cite{Hall:2009bx}. For small $U(1)'$ gauge couplings $\alpha'=g'^2/4\pi$, the DM annihilation processes can be neglected and the DM abundance steadily grows before ``freezing-in'' to a constant value. Conversely, when $\alpha'$ is large, the exchange processes between the DM and $A'$ will be sufficiently important for the whole dark sector to equilibrate at some temperature $T'$ different from the temperature of the SM sector. In this scenario, after the initial production phase, the DM will be able to annihilate into dark photons with a significant rate before the dark sector freezes out. We can therefore identify two main regimes:
\begin{itemize}
    \item[\mylabel{scenarioA}{A.}] \textbf{Pure freeze-in:} in the small $\alpha'$ limit, communication between the DM and $A'$ is too limited to impact the DM number density evolution. The latter therefore depends only on the DM production rate and steadily increases with time until it ``freezes in'' when it can no longer be produced
due to kinematic reasons, i.e. when $T < m_{\chi}$.
\item[\mylabel{scenarioB}{B.}] \textbf{Reannihilation:} when $\alpha'$ is large, $\chi$ and $A'$ are in equilibrium in the early Universe and the dark sector has a temperature $T'$. The DM is first produced through the annihilation of SM fermions and dark photons and later annihilates into dark photons. Hence, the relic density first steadily grows, then diminishes again until the $\chi\chi\rightarrow A'A'$ processes freeze out.
\end{itemize}
The two regimes in fact correspond to two extreme limits of the dark matter-dark photon coupling $\alpha'$. For intermediate values of $\alpha'$ and especially when the dark photon is massive, partial equilibrium can occur. Equilibration is ensured by number-changing processes such as $\chi\,A'\rightarrow \chi \, A'\, A'$ (followed by $A'A'\rightarrow \chi\bar{\chi}$), that can occur either through double Compton scattering or bremsstrahlung. The associated tree-level amplitudes are however infrared divergent, involving large 
logarithms that need to be resummed. Determining to what extent the dark sector is in equilibrium therefore requires particularly involved computations. For studies of dark matter phenomenology, however, the knowledge of the full thermal evolution of the dark sector is often not necessary, and approximate estimates of the final relic density are often enough to derive meaningful theoretical constraints on our models. In this paper, we therefore evaluate the impact of dark sector equilibration in the early Universe on the DM relic density by considering the following two extreme scenarios:
\begin{itemize}
        \item[\mylabel{scenario1a}{1}.] The dark sector is in equilibrium in the early Universe, at a temperature $T'$. For sizable $\alpha'$, dark photon production via inverse decay of SM particles and its subsequent annihilation into pairs of $\chi$ can therefore contribute to the DM production in the early Universe. At later times, the DM can annihilate into pairs of dark photons. Additionally, we assume that the dark photon number density follows its equilibrium value as long as it communicates with the dark matter.
    \item[\mylabel{scenario2b}{2}.] $\chi \bar{\chi} \leftrightarrow A' A'$ equilibrium is not realized. The relic density of $\chi$ particles is hence fully determined by the freeze-in of 
$\bar{f}f \rightarrow \chi \bar\chi$ 
production and, for large $\alpha'$, the freeze-out of $\chi \bar\chi \rightarrow A' A'$ annihilation.
\end{itemize}
The two scenarios correspond to extreme cases of equilibration. 
In the first one, the chemical potentials of the dark matter particle and anti-particle are equal to each other in the early Universe and the chemical potential of the dark photon is always zero. In the second one, the dark sector is always out of equilibrium and the chemical potentials are large. Figure~\ref{fig:1Drelic} shows the evolution of the DM comoving number density $Y_\chi$ in these two scenarios for a given parameter point of our model. For $x = m_{\chi}/T <0.1$, the DM density in scenario~\ref{scenario1a} is much larger  than in scenario~\ref{scenario2b}, since the $A'A'\leftrightarrow\chi\chi$ equilibrium favors DM production. For $x\gtrsim 0.1$, however, this number density sharply drops and closely tracks the one obtained in scenario~\ref{scenario2b}. Ultimately, the resulting relic densities differ by no more than a factor of two, the early Universe dynamics of the dark matter having been almost completely washed out.

In the rest of this work, we compare the values of the DM relic density obtained in scenarios~\ref{scenario1a} and \ref{scenario2b} in order to determine whether the behavior observed in figure~\ref{fig:1Drelic} holds for other choices of parameters. Given our assumptions for these two scenarios, it is safe to assume that {\it if both lead to similar DM relic densities, intermediate scenarios, with partial equilibrium in the dark sector, will lead to similar results}. We first detail the evolution equations corresponding to scenarios~\ref{scenario1a} and~\ref{scenario2b} and discuss interesting limit cases. 

\subsection{Dark sector in equilibrium}
\label{subsec:equilibrium}

In this section we consider the case \ref{scenario1a}, where the dark sector equilibrates at a temperature $T_{\chi} = T_{A'} =T' \neq T$. Additionally, we consider that the dark photon comoving number density always tracks its equilibrium value
\be \label{eq:assumption} Y_{A'}(T_{A'}) \approx Y_{A',eq}(T_{A'}) = \frac{1}{s}
\frac{g_{A'}}{(2\pi)^3} \int_0^\infty  f_{A'}(p,T_{A'})\, d^3 p \ ,\ee 
where $g_{A'} = 3$ is the number of degrees of freedom for a massive spin-1 boson, and $f_{A'}(p,T_{A'})$ is the Bose-Einstein distribution of the $A'$ particles at temperature $T_{A'}$, which is different from the temperature $T$ of the SM sector. Here, $Y_i= n_i/s$ is the ratio of the number density of $n_i$ for a given particle $i$ to the total entropy in the Universe $s$. 
This assumption \eqref{eq:assumption} is valid when $A'$ is much lighter than the dark matter. We therefore expect this scenario to be most relevant to that region of parameter space. 
For the couplings we are interested in, we always have $T'~\ll~T$. We will therefore consider that the dark sector contributions to the total energy density, the Hubble parameter, and the entropy are negligible, and we can write these quantities as
\begin{align}
    \rho &= \rho_{SM} = \frac{\pi^2}{30}g_\text{eff, SM}(T)\,T^4\\
    \label{eq:rho}
    H &= \frac{1}{M_P}\sqrt{\frac{8\pi\rho}{3}}\\
    s &= s_{SM} = \frac{2\pi^2}{45}h_\text{eff, SM}(T)\,T^3 \ ,
\end{align}
where $g_{\rm eff}$ and $h_{\rm eff}$ are the statistical weights for energy and 
entropy, respectively.
In our numerical study we will take the effective numbers of degrees of freedom (as a function of temperature) for the visible sector from the tables in \verb|micrOMEGAs| \cite{Belanger:2018ccd}. 

We compute the dark sector temperature $T'$ following the procedure described in~\cite{Chu:2011be}, using the total energy density of the dark sector $\rho' = \rho_\chi + \rho_{A'}$, defined as
\be
    \rho'(T') = \frac{g_\chi}{2\pi^2}\int_0^\infty \frac{E_\chi\,p^2\,\mathrm{d}p}{e^{E_\chi/T'} + 1} + \frac{g_{A'}}{2\pi^2}\int_0^\infty \frac{E_{A'}\,p^2\,\mathrm{d}p}{e^{E_A'/T'} - 1} \ ,
\ee
where we again impose $\mu_\chi = \mu_{A'} = 0$. Here, since $\chi$ is a Dirac fermion, its number of degrees of freedom is $g_\chi = 4$. Since the dark photon is always massive in our study, $g_{A'} = 3$. For $T' \ll T$, as expected when $\epsilon$ is small, $\rho'$ obeys the following energy transfer equation:
\begin{align}
    \frac{d(\rho'/\rho)}{dT} =& -\frac{1}{H T \rho} \,\sum_f\left[\frac{g_f^2}{32\pi^4}\int ds\,\sigma_{ff\rightarrow\chi\chi}(s)(s - 4 m_f^2)s T K_2\left(\frac{\sqrt{s}}{T}\right)\right.     \nonumber \\
	    & + \left.2\frac{\alpha_{EM}(\epsilon \cos\theta_W q_f)^2}{\pi^2}m_{A'}^4 \left(1+2\frac{m_f^2}{m_{A'}^2}\right)\sqrt{1 - \frac{4m_f^2}{m_{A'}^2}}\,T\,K_2\left(\frac{m_{A'}}{T}\right)\right] \ ,
\end{align}
where we sum over all species of SM fermions $f$ and where the first 
and second terms on the right-hand-side account for energy deposition
into the dark sector via Standard Model annihilation into dark fermions and
inverse decay into dark photons, respectively.

The Boltzmann equations in the non-relativistic limit are
\begin{align}
    \label{eq:boltzmann:full}
    \frac{dY_{\chi}}{dx} &= \frac{s\langle \sigma_{\bar{\chi}\chi\rightarrow \bar{f}f}v\rangle_{_T}}{x H} (Y_{\chi,eq}^2(T) - Y_\chi^2) + \frac{s\langle \sigma_{\bar{\chi}\chi\rightarrow A'A'}v\rangle_{_{T'}}}{x H} \left(\frac{Y_{\chi,eq}^2(T')}{Y_{A',eq}^2(T')}Y_{A'}^2 - Y_\chi^2\right)     \nonumber \\
    \frac{dY_{A'}}{dx} &= \frac{\Gamma_{A'\rightarrow \bar{f}f}}{x H} (Y_{A',eq}(T) - Y_{A'}) - \frac{s\langle \sigma_{\bar{\chi}\chi\rightarrow A'A'}v\rangle_{_{T'}}}{x H} \left(\frac{Y_{\chi,eq}^2(T')}{Y_{A',eq}^2(T')}Y_{A'}^2 - Y_\chi^2\right).
\end{align}
The subscript for the velocity-averaged cross sections indicates the temperature at which they should be evaluated. Note, in particular, that the $\chi\chi \rightarrow A'A'$ cross section is evaluated at $T'$ and not at $T$.
%%%%
For small $\epsilon$, we can approximate these equations by 
\begin{align}
    \label{eq:boltzmann:approx}
    \frac{dY_{\chi}}{dx} &\approx \frac{s\langle \sigma_{\bar{\chi}\chi\rightarrow \bar{f}f}v\rangle_{_T}}{x H} Y_{\chi,eq}^2(T) + \frac{s\langle \sigma_{\bar{\chi}\chi\rightarrow A'A'}v\rangle_{_{T'}}}{x H} \left(\frac{Y_{\chi,eq}^2(T')}{Y_{A',eq}^2(T')}Y_{A'}^2 - Y_\chi^2\right)  \nonumber \\
    \frac{dY_{A'}}{dx} &\approx \frac{\Gamma_{A'\rightarrow \bar{f}f}}{x H} Y_{A',eq}(T) - \frac{s\langle \sigma_{\bar{\chi}\chi\rightarrow A'A'}v\rangle_{_{T'}}}{x H} \left(\frac{Y_{\chi,eq}^2(T')}{Y_{A',eq}^2(T')}Y_{A'}^2 - Y_\chi^2\right).
\end{align}
Assuming that $A'$ is fully thermalized, we can finally write the dark matter evolution equation as
\begin{align}
    \label{eq:boltzmann:approx1}
    \frac{dY_{\chi}}{dx} &\approx \frac{s\langle \sigma_{\bar{\chi}\chi\rightarrow \bar{f}f}v\rangle_{_T}}{x H} Y_{\chi,eq}^2(T) + \frac{s\langle \sigma_{\bar{\chi}\chi\rightarrow A'A'}v\rangle_{_{T'}}}{x H} \left(Y_{\chi,eq}^2(T') - Y_\chi^2\right).
\end{align}
When $\alpha'$ is large, the last term in this equation dominates over the $f\bar{f}\rightarrow \chi\chi$ production term for an extended period of time in the early Universe and the DM density $Y_\chi$ is nearly equal to its equilibrium value $Y_{\chi,eq}(T')$. As detailed in~\cite{Chu:2011be}, the dark matter-dark photon exchange processes will ultimately freeze out, 
when the $A'A'\rightarrow \chi\chi$ production term shuts off for 
$T' \lesssim m_{\chi}$. This will lead to the DM density either stabilizing to its final value or evolving further following the equation
\begin{align}
    \label{eq:boltzmann:approx2}
    \frac{dY_{\chi}}{dx} &\approx \frac{s\langle \sigma_{\bar{\chi}\chi\rightarrow \bar{f}f}v\rangle_{_T}}{x H} Y_{\chi,eq}^2(T) - \frac{s\langle \sigma_{\bar{\chi}\chi\rightarrow A'A'}v\rangle_{_{T'}}}{x H}  Y_\chi^2.
\end{align}
In practice, the $f\bar{f}\rightarrow \chi\chi$ production term and the $\chi\chi\rightarrow A'A'$ annihilation term will often balance each other, leading to a new temporary equilibrium, dubbed the ``Quasi-Static Equilibrium'' (QSE) in \cite{Chu:2011be}.
In this case, the relic density of the DM is approximately given by
\begin{align}
    \label{eq:yqse}
	Y_\chi \approx \sqrt{\frac{\langle\sigma_{\bar{\chi}\chi\rightarrow \bar{f}f}v\rangle_{_T}}{\langle \sigma_{\bar{\chi}\chi\rightarrow A'A'}v\rangle_{_{ T'}}}}\, Y_{\chi, eq}(T)\equiv Y_{QSE}(T)
\end{align}
before freeze-out. The simplified scenario in this paragraph is therefore fully defined by the analytic expression \eqref{eq:yqse} and the freeze-out temperature. A quasi-static equilibrium behavior can for example be observed in figure~\ref{fig:1Drelic} for $0.1\lesssim x \lesssim 10$.

\subsection{Fully out-of-equilibrium scenario}
\label{sec:fos}

In the simplified scenario \ref{scenario2b}, we assume that the DM and the dark photon never equilibrate and, in particular, that dark matter production from dark photon annihilation can be neglected. The dark matter number density will therefore be set first by DM production from SM fermion annihilation and later, when the $U(1)'$ structure constant $\alpha'$ is large, by DM annihilation into dark photons.
In this regime, the Boltzmann equation for $\chi$ in the non-relativistic limit is given by
\bea 
    \label{eq:boltzmann:reann}
    \frac{dY_{\chi}}{dx} = \frac{s\langle \sigma_{\bar{\chi}\chi\rightarrow \bar{f}f}v\rangle_{_T}}{x H} Y_{\chi,eq}^2(T) - \frac{s\langle \sigma_{\bar{\chi}\chi\rightarrow A'A'}v\rangle_{_{T'}}}{x H}  Y_\chi^2 \ .
 \eea
As before, we assumed that $Y_\chi(T) \ll Y_{\chi,eq}(T)$, since the $\epsilon$ coupling connecting the DM and the SM is very small. Since the DM annihilation into dark photons becomes important only at late times and therefore low temperatures, and occurs dominantly in the $s$-wave, we can safely approximate the $\overline{\chi}\chi \rightarrow A'A'$ cross-section by its value at zero temperature, i.e.
\be
\langle \sigma_{\bar{\chi}\chi\rightarrow A'A'}v\rangle_{_{T'}} \approx \langle \sigma_{\bar{\chi}\chi\rightarrow A'A'}v\rangle_{_0}.
\ee
The final simplified Boltzmann equation then is
 \bea
    \label{eq:boltzmann:reannapprox}
    \frac{dY_{\chi}}{dx} = \frac{s\langle \sigma_{\bar{\chi}\chi\rightarrow \bar{f}f}v\rangle_{_T}}{x H} Y_{\chi,eq}^2(T)  - \frac{s\langle \sigma_{\bar{\chi}\chi\rightarrow A'A'}v\rangle_{_0}}{x H}  Y_\chi^2 \ .
 \eea

 When $\alpha'$ is small, this equation can be approximated by a ``pure freeze-in'' equation of the form
 \bea
    \label{eq:boltzmann:freezinsimp}
    \frac{dY_{\chi}}{dx} = \frac{s\langle \sigma_{\bar{\chi}\chi\rightarrow \bar{f}f}v\rangle_{_T}}{x H} Y_{\chi,eq}^2(T) \ .
 \eea
 In this regime the interactions $A'A' \leftrightarrow \bar{\chi}\chi$ play no significant role in determining the relic density of either dark sector particle. The simplified evolution equation can be solved analytically, taking the late time limit. The result is given by \cite{Hall:2009bx}
 \be Y(t\rightarrow \infty) \approx \frac{1215 \sqrt{\frac{5}{2}} e^4  M_p \epsilon ^2}{512 \pi^6 \sqrt{g_{eff}} h_{eff} m_\chi}  \ . \ee

 Conversely, when $\alpha'$ is large, the last term of equation~\ref{eq:boltzmann:reannapprox} can balance out the DM production term. In this case, just as in the equilibrated case discussed in section~\ref{subsec:equilibrium}, the DM reaches a quasi-static equilibrium until the $f\bar{f}\rightarrow \chi\chi$ production term shuts off and the annihilation $\bar{\chi}\chi\rightarrow A'A'$ process freezes out. The associated relic density will also be given by equation~\ref{eq:yqse}, evaluated at the freeze-out temperature $T_f$ and taking $T' = 0$. For parameter points where a QSE occurs, we can therefore already predict that the only discrepancy in the final DM relic density between scenarios~\ref{scenario1a} and~\ref{scenario2b} will arise from computing the $A'A'\rightarrow\chi\chi$ cross-sections at $T'$ and zero respectively. As can be seen in figure~\ref{fig:1Drelic}, this difference is not expected to be significant since temperatures in the dark sector are usually low at late times. Finally, for intermediate $\alpha'$ values, the equation~\ref{eq:boltzmann:reannapprox} has to be solved numerically but this procedure is particularly simple as it does not involve computing chemical potentials and temperatures in the dark sector.

In the following section, we will perform a scan over the different parameters of our model to compute the DM relic densities in the equilibrated scenario~\ref{scenario1a} and the out-of-equilibrium scenario~\ref{scenario2b} in order to assess the impact of the internal dark sector dynamics in the early Universe on the final DM relic density.

\section{Numerical scan and results}
\label{sec:numerical}
In this section, we determine in what region of the parameter space the dark matter relic density is nearly independent of the internal dark sector dynamics at high temperatures. To this end, we perform a numerical scan over the different parameters of our model and compare the relic density obtained in the fully out-of-equilibrium scenario~\ref{scenario2b}, via integration of Eq.~\ref{eq:boltzmann:reannapprox}, to the one obtained in scenario~\ref{scenario1a} where either $A'$ or the whole dark sector is fully equilibrated, via integration of Eq.~\ref{eq:boltzmann:approx1}.
The true relic density is expected to lie in between those limiting cases.
Details on how we evaluate the relic densities numerically are given in 
Appendix B.

Our model involves four parameters, $\epsilon$, $\alpha_\mathrm{DM}$, $m_\mathrm{DM}$, and $m_{A'}$, that can vary over orders of magnitude. We therfore perform a scan over the following ranges:
\begin{eqnarray}
    &m_\mathrm{DM} \in [0.2~\mathrm{GeV}, 1~\mathrm{TeV}],&\text{step size }0.1 \nonumber \\
    &m_{A'} \in [2~\mathrm{MeV},  \min\{m_\mathrm{DM}, 1~\mathrm{GeV} \}],&\text{step size }0.5 \nonumber \\
    &\epsilon \in [10^{-15}, 10^{-9}],&\text{step size }0.5\nonumber \\
    &\alpha_\mathrm{DM} \in [10^{-8}, 10^{-2}],&\text{step size }0.5.
\end{eqnarray}
 The dimensionless step sizes above are logarithmic, so, for a given step size $\delta$, the value of a parameter $p$ with minimum value $p_\mathrm{min}$ at step $n$ will be
\be
    p = p_\mathrm{min} \times 10^{n\delta}.
\ee
We express the discrepancy between the (partially) equilibrated scenario~\ref{scenario1a} and the out-of-equilibrium scenario~\ref{scenario2b} by
\begin{equation}
    \mathcal{E} = \frac{\Omega_\mathrm{scenario2}}{\Omega_\mathrm{scenario1}}.
\end{equation}

\begin{figure}[h]
    \centering
    \begin{subfigure}[b]{0.47\textwidth}
    \includegraphics[width=\textwidth]{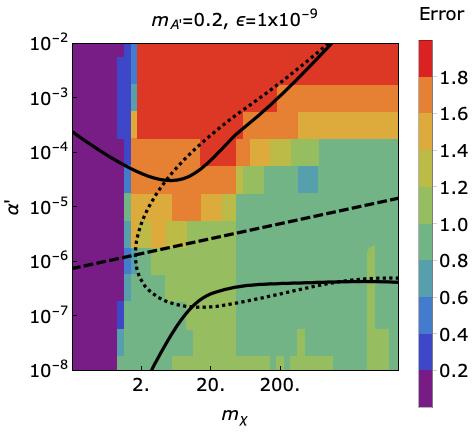}
%     \caption[t]{ $m_{A'} = 2 \times 10^{-1}$, $\epsilon =  10^{-9}$}
          \vspace{0.1cm}
    \end{subfigure} 
    \hspace{0.5cm}
    \begin{subfigure}[b]{0.47\textwidth}
    \includegraphics[width=\textwidth]{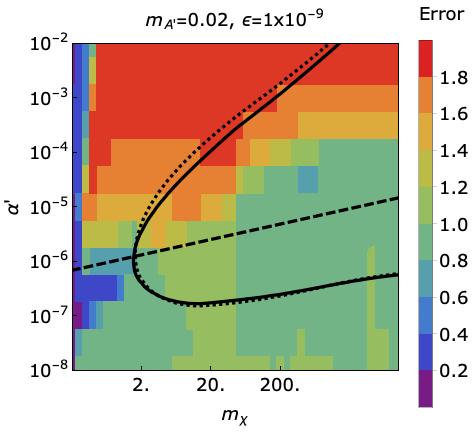}
%     \caption{ $m_{A'} = 2 \times 10^{-2}$, $\epsilon =  10^{-9}$}
         \vspace{0.1cm}
     \end{subfigure} 
    \begin{subfigure}[b]{0.47\textwidth}
    \includegraphics[width=\textwidth]{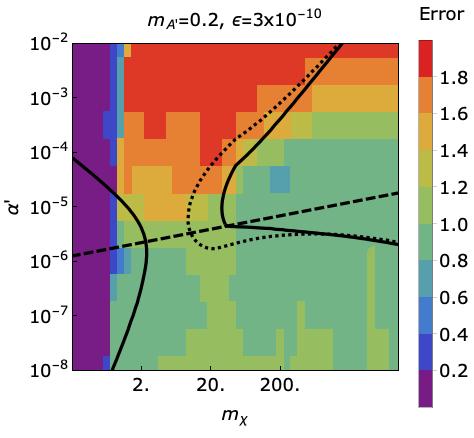}
   %  \caption{ $m_{A'} = 2 \times 10^{-1}$, $\epsilon = 3 \times 10^{-10}$}
    \end{subfigure}
     \hspace{0.5cm} 
    \begin{subfigure}[b]{0.47\textwidth}
    \includegraphics[width=\textwidth]{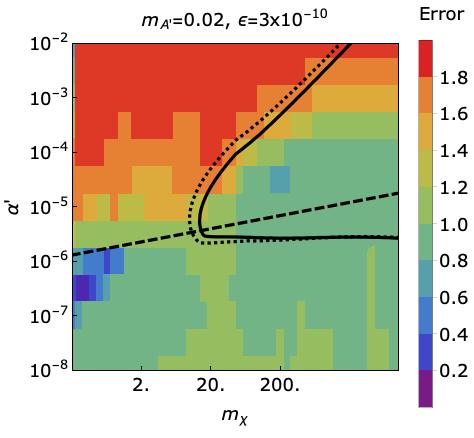}
%    \caption{ $m_{A'} = 2 \times 10^{-2}$, $\epsilon = 3 \times 10^{-10}$}
    \end{subfigure} 
    \caption{Relative error  $\mathcal{E}$ versus $\alpha'$ and $m_\chi$, where masses are in GeV. We study values of $\epsilon$ that are small enough that the dark sector is out of thermal equilibrium with the visible sector, but large enough to produce the observed relic abundance.  The dashed line is the approximate transition between the freeze-in regime (below the line) and the re-annihilation regime (above the line), the solid line indicates where the relic abundance calculated using the equilibrium calculation (scenario~\ref{scenario1a})  matches the observed relic abundance, and the dotted line indicates where the relic abundance calculated using the simplified calculation (scenario~\ref{scenario2b}) matches the observed abundance.}\label{fig:scan}
\end{figure}

Fig. \ref{fig:scan} shows the color-coded values of $\mathcal{E}$ for our parameter scan in the $(\alpha', m_\chi)$ space for fixed values of the portal coupling $\epsilon$ and the mediator mass $m_{A'}$. The black line in each panel shows the approximate transition region between pure freeze-in (regime~\ref{scenarioA}) and re-annihilation (regime~\ref{scenarioB}).
Here, we chose values of $\epsilon$ sufficiently large to obtain the observed DM relic density in some regions of the parameter space. We note that, although we chose to show our results only for specific values of $\epsilon$ and $m_{A'}$ for the sake of clarity, the behaviors and numerical results shown here are representative of the ones we obtained in the rest of the parameter space.

From Fig.~\ref{fig:scan} we first observe that, in most of the parameter space, assuming the out-of-equilibrium scenario~\ref{scenario2b} leads to values of the DM relic density that are within a factor of two of the ones obtained in the equilibrated scenario~\ref{scenario1a}. 
The only region in which scenario~\ref{scenario2b} significantly differs from scenario~\ref{scenario1a} is the $m_{A'} > 10^{-1}\,m_\chi$ region, where since $A'$ is heavy, equilibrium is unlikely to occur and therefore assuming the fully equilibrated scenario~\ref{scenario1a} leads to overestimating the relic density. Note that, while the relic density contours corresponding to the PLANCK value coincide in scenarios~\ref{scenario1a} and \ref{scenario2b} for $m_\mathrm{DM}\gg m_\mathrm{m_{A'}}$, they start strongly diverging as the dark matter mass becomes lower. The $m_\mathrm{DM}\sim m_\mathrm{A'}$ region therefore deserves a more in-depth treatment.

Since the abundances predicted in scenarios~\ref{scenario1a} and \ref{scenario2b} bracket the true relic abundance we can conclude the following: as long as the dark photon is much lighter than the dark matter, the degree of equilibration in the dark sector has an extremely limited influence on the value of the DM relic density. It is therefore possible to considerably simplify the treatment of a wide range of dark matter models with a light vector mediator.  At first sight, our results may appear very surprising. We have argued in the
introduction that thermalization uncertainties in the considered dark matter scenario may potentially indroduce orders-of-magnitude uncertainties in the determination of relic abundances. However, the closeness of predicted abundances in scenarios~\ref{scenario1a} and \ref{scenario2b} maybe understood from the following observation. These two scenarios distinguish themselves by the
$A'A'\rightarrow \chi\chi$ production term and the evaluation of the 
$\chi\chi\rightarrow A'A'$ cross section at either $T'\neq 0$ or $T'=0$. We have already noted that the latter difference is minimal for $s$-wave annihilation
at $\chi$ freeze-out. Dark matter $\chi$ production via $A'$ self-annihilation may become important for large $A'$ densities (i.e. large $\epsilon$ and large $T'$)
as well as large $A'$ self-annihilation cross section (i.e. large $\alpha'$). In this case one could expect for scenario 1 to give a much larger abundance than
scenario 2. However, large $\epsilon$ and large $\alpha'$ also implies the
likely re-annihilation of the produced dark matter, with the end effect that
though in the early Universe there may be orders of magnitude differences in
the $\chi$ abundance (cf. Fig 1), the freeze-out value is essentially the same.
It is not clear whether our results are generic to other dark sector
Lagrangians; we leave this to future work.

\section{Conclusion and discussion}
\label{sec:discussion}
In this work, we have demonstrated that, in many cases, the complicated intermediate stages for the production of dark matter via the freeze-in mechanism can be skipped in the calculation of the final relic abundance.  This implies, for the great majority of parameter space,  the relic abundance can be approximated well through an analytic or a simplified numeric calculation. This is a great simplification and in particular circumvents the great uncertainties associated with the relic abundance calculation, in particular from number-changing processes. Specifically, we demonstrated the close resemblance of the fully equilibrated and simplified scenarios with a direct numerical comparison, shown in Fig.~\ref{fig:scan}. It is seen that the approximation holds best for light mediators, i.e. large mass gap $m_\chi/m_{A'}$.

It is interesting to note that for part of the parameter space considered in this work, the relic abundance is the only distinguishing feature of the dark sector. Direct detection constraints are weakened for light mediators~\cite{Kaplinghat:2013yxa}, and indirect detection constraints are weakened as the dominant annihilation channel is into dark mediators~\cite{Elor:2015bho}. The mediator mixing $\epsilon$ is generally too large in the scenarios considered here to be probed by CMB or BBN constraints, with the mediator decaying just before nucleosynthesis starts. Constraints due to decays during BBN exist in the part of parameter space with smaller mixing~\cite{Berger:2016vxi}. Assuming no hidden sector decays, dark photons that are lighter than the electron mass are stable, dark matter matter candidates~\cite{Nelson:2011sf}. Other constraints do not apply in the parameter space considered here as the gauge boson mass is too large. For example, strong constraints on the massless $A'$ scenario come from milli-charged particles, and from the effects of long-range interactions on DM haloes (see for example~\cite{Robertson:2016xjh,Knapen:2017xzo}). Such interactions would deform the DM halo profiles in elliptic galaxies, which gives a strong constraint which essentially rules out the re-annihilation scenario in this case.
%
%Massless or very light dark photons with small mixing are also senstive to a BBN constraint from effective neutrino number. \textcolor{blue}{JB: not relevant for the models considered here--decays too early.  Remove?} The constraint is not very strong, but sets an upper bound on the value of the temperature of the dark sector. 

\section*{Acknowledgements}
The authors thank C. Sun, R. Caldwell and C. Smith for valuable discussions. 
This work was performed in part at the Aspen Center for Physics, which is supported by National Science Foundation grant PHY-1607611. The work of JB is supported in part by U.S. Department of Energy grant no. de-sc0007914 and in part by PITT PACC. TRIUMF receives federal funding via a contribution agreement with the National Research Council of Canada and the Natural Science and Engineering Research Council of Canada. DGEW is supported by a Burke faculty fellowship. SEH has been supported by the NWO Vidi grant ``Self-interacting asymmetric dark matter''. 

\appendix
\section{Velocity-averaged cross sections}\label{sec:crosssections}
Here, we follow the procedure described in \cite{Edsjo:1997bg} to calculate the velocity-averaged cross sections needed to evaluate the condition in equation~\ref{eq:equilibrium}. We consider $W_{ff} = \left[\sigma s v\right]_{\bar{\chi}\chi\rightarrow \bar{f}f}$, defined for the photon channel as:
\begin{align}
    \left[\sigma s v\right]_{\bar{\chi}\chi\rightarrow \bar{f}f}[s, m_\chi, m_f] &= N^c_f\frac{4\pi \alpha_{EM}^2 \epsilon^2 \alpha'^2 q_f^2}{3s^2}\sqrt{s - 4m_f^2} (s + 2 m_f^2)(s + 2 m_\chi^2) \ ,
\end{align}
Now, using equation (65) from \cite{Edsjo:1997bg}, the velocity-averaged cross section can be written as
\begin{align}
    \langle\sigma_\text{eff}v\rangle = \frac{\int_{p_\text{min,f}}^\infty \mathrm{d}p\, p^2 W(p,m)K_1(\sqrt{s}/T)}{m^4 T K_2^2(x)} \ ,
\end{align}
with $x = m/T$ and $p_{\text{min},f} = \sqrt{\text{max}\left\{0, m^2 - m_f^2\right\}}$. For $x < 1$, we define our integration variable as $z = p/T$, which gives
\begin{align}
    \langle\sigma_\text{eff}v\rangle = T^2\frac{\int_{z_\text{min,f}}^\infty \mathrm{d}z\, z^2 W(z, x)K_1(2\sqrt{z^2 + x^2})}{m^4 K_2^2(x)} \ .
\end{align}
To properly compute the Boltzmann equations, it makes more sense to define directly $\langle\sigma_\text{eff}v\rangle T^2$, which is equal to
\begin{align}
    \langle\sigma_\text{eff}v\rangle T^2 = \frac{\int_{z_\text{min,f}}^\infty \mathrm{d}z\, z^2 W(z, x)K_1(2\sqrt{z^2 + x^2})}{x^4 K_2^2(x)} \ .
\end{align}
In particular, when $x$ goes to $0$, we have
\begin{align}
    \langle\sigma_\text{eff}v\rangle T^2 = \frac{1}{4}\int_{0}^\infty \mathrm{d}z\, z^2 W(z, 0)K_1(2z)  \ .
\end{align}
In practice, in the code we choose the upper limit of integration to be $z_\text{max} = z_\text{min,f} + 10$ (so a different one for each fermion species).

Now, when $x > 1$, we use $z = p/m$ as an integration variable and we obtain
\begin{align}
    \langle\sigma_\text{eff}v\rangle T^2 = \frac{\int_{z_\text{min,f}}^\infty \mathrm{d}z\, z^2 W(z, 1)K_1(2x\sqrt{z^2 + 1})}{x K_2^2(x)} \ .
\end{align}
We use a similar approach to evaluate the velocity-averaged cross section for $\chi\chi \rightarrow A'A'$.

\section{Details on the numerical evaluation of relic abundances}
As the equations to solve numerically may be quite stiff, we evaluate them in
the following way.
For each parameter point, we estimate how far the dark sector is from equilibrium. The equilibrium condition, derived from 
Eq.~\ref{eq:boltzmann:approx1} with ${\rm d} Y_{\chi}/{\rm d} x
\to {\rm d} Y_{\chi,eq}(T')/dx$, can be written as
\begin{align}
    \left|\frac{Y_{\chi}^2}{Y_{\chi, eq}^2(T')}  - 1\right|\approx \left|\frac{\frac{s\langle \sigma_{\chi\chi\rightarrow ff}v\rangle_{_T}}{x H} Y_{\chi,eq}^2(T) - \frac{dY_{\chi, eq}(T')}{dx}}{Y_{\chi, eq}^2(T')\frac{s\langle \sigma_{\chi\chi\rightarrow A'A'}v\rangle_{_{T'}}}{x H}}\right| \  \ll 1.
    \label{eq:equilibriumcond}
\end{align}
We consider that the parameter points that verify this condition are also
in the reannihilation regime. In order to simplify our analysis, instead of directly integrating equation~\ref{eq:boltzmann:approx1}, we first check whether the quasi-static equilibrium (QSE) mentioned in section~\ref{subsec:equilibrium} is reached. For this, we use a condition analogous to equation~\ref{eq:equilibriumcond}, replacing $Y_{\chi, eq}(T')$ by $Y_{QSE}(T)$. 
If this QSE is reached, then the late time evolution of the dark matter density will depend on its early Universe dynamics only through the value of the dark sector temperature at the time when the $A'A'\leftrightarrow \chi\chi$ processes freeze out. Thus, the discrepancy between the values of the DM relic density in scenarios~\ref{scenario1a} and \ref{scenario2b} would entirely stem from evaluating the $\chi\chi\rightarrow A'A'$ cross-section at different temperatures in the two regimes. Given the weak sensitivity of said cross-section in the temperature at late times, this discrepancy is expected to remain extremely moderate. In our numerical scan, we therefore consider the following three configurations:
\begin{itemize}
    \item \emph{Equilibrium within the dark sector \textbf{and} QSE:} here, the DM production from $A'$ annihilation will only weakly affect the value of the DM relic density, as explained above. This case can therefore be well described by the simplified scenario~\ref{scenario2b} from the introduction. To quantify the (small) error arising from the $T' = 0$ approximation, we integrate the Boltzmann equations~\ref{eq:boltzmann:approx1} and \ref{eq:boltzmann:reannapprox} starting from the time of departure from QSE and compare the resulting relic densities $\Omega_\mathrm{scenario 1}$ and $\Omega_{\rm scenario 2}$.
    \item \emph{Equilibrium within the dark sector \textbf{without} QSE:} here, the DM production from $A'$ annihilation could potentially sizably affect the DM relic density. In order to determine whether scenario~\ref{scenario2b} is still a reasonable approximation, we integrate the Boltzmann equations~\ref{eq:boltzmann:approx1} and \ref{eq:boltzmann:reannapprox} starting from the dark sector freeze-out time and compare the resulting relic densities $\Omega_\mathrm{scenario1}$ and $\Omega_{\rm scenario2}$.
    \item \emph{Dark sector out of equilibrium (freeze-in):} here, we compare the relic densities $\Omega_\mathrm{scenario 1}$ and $\Omega_\mathrm{scenario 2}$ obtained by solving the Boltzmann equations associated with scenario~\ref{scenario1a} (equation~\ref{eq:boltzmann:approx1}) and scenario~\ref{scenario2b} (equation~\ref{eq:boltzmann:reannapprox}) respectively. Note that in the latter case, the contribution of the term proportional to $\alpha'$ will often be negligible.    
\end{itemize}

% \bibliography{mono}{}
 \providecommand{\href}[2]{#2}\begingroup\raggedright\endgroup

\end{document}